\documentclass[12pt,axodraw]{JHEP3}
\usepackage[english]{babel}
\usepackage{amssymb}
\usepackage{amsfonts}
\usepackage{amsmath}
\usepackage{graphicx}
\usepackage{subfigure}
\usepackage{epsfig}
\usepackage{cite}
\usepackage{psfrag}
\usepackage{slashed}
\usepackage{axodraw}

\usepackage[normalem]%
{ulem}  
\newcommand{\old}[1]{}

\newcommand{\cA}{{\cal A}}  
  
  \newcommand{\cF}{{\cal F}}

  \newcommand{\cV}{{\cal V}}

\newcommand{\bbA}{{\mathbb A}}  \newcommand{\bbB}{{\mathbb B}}

\newcommand{\be}{\begin{equation}} \newcommand{\ee}{\end{equation}}
\newcommand{\bea}{\begin{eqnarray}} \newcommand{\eea}{\end{eqnarray}}
\newcommand{\beann}{\begin{eqnarray*}}  \newcommand{\eeann}{\end{eqnarray*}}
\newcommand{\bfig}{\begin{figure}} \newcommand{\efig}{\end{figure}}
\newcommand{\ba}{\begin{array}} \newcommand{\ea}{\end{array}}
\newcommand{\bcen}{\begin{center}} \newcommand{\ecen}{\end{center}}
\newcommand{\btab}{\begin{tabular}} \newcommand{\etab}{\end{tabular}}

\def\tr{\operatorname{tr\:}}

\newcommand{\mubar}{\bar{\mu}}
\newcommand{\ms}{\overline{\mathrm{MS}}}
\newtheorem{Proposition}{Proposition}[section]

\newtheorem{Theorem}{Theorem}[section]
\newtheorem{Lemma}{Lemma}[section]
\newtheorem{Corrolary}{Corrolary}[section]

\newcommand{\bp}{\begin{Proposition}}	\newcommand{\ep}{\end{Proposition}}
\newcommand{\bt}{\begin{Theorem}}	\newcommand{\et}{\end{Theorem}}
\newcommand{\bl}{\begin{Lemma}}		\newcommand{\el}{\end{Lemma}}
\newcommand{\bc}{\begin{Corrolary}}	\newcommand{\ec}{\end{Corrolary}}

\title{
Holographic Anomalous Conductivities and the Chiral Magnetic Effect
}
\author{ Antti Gynther$^{a}$, Karl Landsteiner$^{b}$,  Francisco Pena-Benitez$^{b}$, Anton Rebhan$^{a}$\\
$^{a}$Institut f\"ur Theoretische Physik\\
Technische Universit\"at Wien\\
Wiedner Hauptstrasse 8-10, 1040 Vienna, Austria\\
E-mail: \email{gynthera@hep.itp.tuwien.ac.at}, \email{rebhana@hep.itp.tuwien.ac.at}\\
\\
$^{b}$Instituto de F\'\i sica Te\' orica CSIC-UAM\\
C-XVI Universidad Aut\' onoma de Madrid\\
E-28049  Madrid, Spain\\
E-mail: \email{karl.landsteiner@uam.es, fran.penna@uam.es}
}

\keywords{Gauge-gravity correspondence, QCD, Chiral Lagrangians}
\preprint{IFT-UAM/CSIC-10-35}

\abstract{We calculate anomaly induced conductivities from a holographic gauge theory model using Kubo formulas,
making a clear conceptual distinction between thermodynamic state variables such as chemical potentials and external
background fields. This allows us to pinpoint ambiguities in previous holographic calculations of the chiral magnetic
conductivity. We also calculate the corresponding anomalous current three-point functions in special kinematic regimes.
We compare the holographic results to weak coupling calculations using both dimensional regularization and cutoff
regularization. In order to reproduce the weak coupling results it is necessary to allow for singular holographic gauge field configurations
when a chiral chemical potential is introduced for a chiral charge defined through a gauge invariant but non-conserved chiral density. We argue that this is appropriate for actually addressing charge separation due to the chiral magnetic effect.
}

\begin{document}
\section{Introduction}

The chiral anomaly of QED is responsible for two particularly interesting effects of
strong magnetic fields in dense, strongly interacting matter as found in
neutron stars
or heavy-ion collisions. At large quark chemical potential $\mu$, chirally
restored quark matter gives rise to an axial current parallel to the magnetic
field \cite{Son:2004tq,Metlitski:2005pr,Newman:2005as}
\be\label{J5}
\mathbf J_5=\frac{eN_c}{2\pi^2}\mu \,\mathbf B,
\ee
which may lead to observable effects in strongly magnetized neutron stars
\cite{Charbonneau:2009ax}.

In heavy-ion collisions, one expects initial magnetic fields that momentarily
exceed even those found in magnetars. It has been proposed by Kharzeev et al.\
\cite{Kharzeev:2004ey,Kharzeev:2007tn,Kharzeev:2007jp,Fukushima:2008xe,Kharzeev:2009pj}
that the analogous effect \cite{Alekseev:1998ds}
\be\label{CME}
\mathbf J=\frac{e^2N_c}{2\pi^2}\mu_5 \,\mathbf B,
\ee
where $\mathbf J$ is the electromagnetic current and $\mu_5$ a chemical
potential for an asymmetry in the number of right and left chiral quarks, could
render observable event-by-event P and CP violations from topologically
nontrivial gluon configurations. Indeed, there is recent experimental evidence
for this ``chiral magnetic effect'' (CME)
in the form of 
charge separation in heavy ion collisions with respect to the reaction plane
\cite{Abelev:2009uh,Voloshin:2008jx} (see however \cite{Wang:2009kd,Asakawa:2010bu}),
whose normal vector is expected to coincide with the direction of strong
initial magnetic fields. For lattice studies of the effect, see for example \cite{Buividovich:2009wi,Buividovich:2010tn},

The anomalous conductivities (\ref{J5}) and (\ref{CME})
have recently also been studied in holographic models
of QCD by introducing chemical potentials for left and right chiral quarks
as boundary values for corresponding bulk gauge fields \cite{Lifschytz:2009si,Yee:2009vw}. However, it was pointed out by Ref.~\cite{Rebhan:2009vc} 
that in these calculations
the axial anomaly was not realized in covariant form and that the corresponding
electromagnetic current was not strictly conserved. Correcting the situation
by means of Bardeen's counter-term \cite{Bardeen:1969md,Hill:2006ei}
instead led to a vanishing\footnote{It has been argued that this is an artifact of the grand canonical ensemble and that the weak coupling result would be recovered in the canonical ensemble \cite{YeeBNL}.} result for
the electromagnetic current in the holographic QCD model
due to Sakai and Sugimoto \cite{Sakai:2004cn,Sakai:2005yt}\footnote{In Ref.~\cite{Gorsky:2010xu} a finite
result was obtained in a bottom-up model that is nonzero only
due to extra scalar fields.}, while recovering the result (\ref{J5}) for
the anomalous axial conductivity.

Indeed, the two anomalous conductivities (\ref{J5}) and (\ref{CME}) differ
in that in the former case there is no difficulty with introducing a chemical
potential for quark number, while a chemical potential for chirality refers
to a chiral current that is either gauge invariant and anomalous or
conserved but not gauge invariant. In Ref.~\cite{Alekseev:1998ds},
the chiral magnetic effect (\ref{CME}) was shown to be an exact result
when the chiral chemical potential is conjugate to conserved chiral charges that
are, however,
only gauge invariant when integrated over all of space in spatially homogenous
situations (this point was most recently
also made in Ref.~\cite{Rubakov:2010qi}). 

However, charge separation in heavy-ion collisions clearly
calls for inhomogeneous situations, since with $\mathbf\nabla\cdot\mathbf B\equiv 0$
we have
\be
\frac{\partial\rho}{\partial t}
=-\mathbf\nabla\cdot\mathbf J=-\frac{e^2N_c}{2\pi^2}\mathbf B\cdot
\mathbf\nabla \mu_5.
\ee
It would therefore appear important to introduce a chiral chemical
potential conjugate to gauge invariant axial currents, despite them being 
anomalous. 
At least
as long as the electric field is zero and as long as the chiral charge
decay rate is suppressed (as it is in the large-$N_c$ limit \cite{Moore:2010jd}, and the fact that we indeed find time-independent solutions in the presence of a chiral chemical potential is a reflection of this fact), 
it should be admissible to consider chemical potentials defined 
with respect to the gauge invariant chiral density in
a thermodynamic description.
Such a chiral chemical potential thus serves as a model parameter
for the imbalance between the number of left-handed and right-handed
fermions that is assumed to be induced by topologically non-trivial gluon field configurations during the out-of-equilibrium early stages of a heavy ion collision.

It is however
important
to distinguish between thermodynamic state variables such as chemical potentials
and background gauge fields (as also pointed out
by  Ref.\ \cite{Rubakov:2010qi}). 
{The holographic dictionary instructs us to construct a functional
of boundary fields and $n$-point functions are obtained by functional differentiation
with respect to the boundary fields. For a gauge field, the expansion close to the 
boundary takes the form}
$$
A_\mu(x,r) = A_\mu^{(0)}(x) + \frac{ A_\mu^{(2)}(x)}{r^2} + \dots\,.
$$
{The leading term in this expansion is the source for the current $J^\mu$.
The sub-leading term is often identified with the one-point function of the current.
This is, however, not true in general. As has been pointed out in Ref.\ \cite{Rebhan:2009vc},
in the presence of a bulk Chern-Simons term, the current receives also contributions
from the Chern-Simons term and $A_\mu^{(2)}(x)$ can, in general, not be identified with the vev 
of the current. On the other hand, a constant value of $A_0^{(0)}$ is often identified with a chemical 
potential. This is, however, slightly misleading since the holographic realization of the chemical potential
is given by the potential difference between the boundary and the horizon \cite{Ghoroku:2007re} and only in
a gauge in which $A_0$ vanishes at the horizon such an identification can be made. Even in this
case, we have to keep in mind that the boundary value of the gauge field is the source of the current
whereas the potential difference between horizon and boundary is the chemical potential. 
We will keep this distinction explicit in this paper.}

In section 2 we shall show that by {distinguishing the chemical potential from the background gauge fields}
one can reproduce the usual result (\ref{CME}) for the chiral magnetic effect
when $\mu_5$ refers to the gauge invariant chiral current\footnote{The chiral current is
gauge invariant under the non-anomalous vectorial gauge transformations.} and for
a strictly conserved electromagnetic current $J$. However, this
requires singular gauge field configurations in the bulk of AdS space, which
appears to be the price to pay for having introduced a chemical potential
for an anomalous charge.
We also reproduce the uncontested result (\ref{J5}) for the axial current
at finite quark chemical potential and magnetic fields, as well as a new
anomalous conductivity, albeit one of perhaps 
mere academic interest as it refers
to nonzero axial magnetic fields. Moreover, we derive results for anomalous
three-point functions in certain kinematic limits. In section 3 
we reproduce all these results
in weak coupling calculations using gauge invariant
dimensional regularization without and cutoff regularization with the need
for introducing Bardeen's counter-term.

\section{Holographic Kubo formulas for anomalous conductivities}

We will consider the simplest possible holographic model for one quark flavor in a chirally restored deconfined phase.\footnote{The even simpler model considered in
Ref.~\cite{Rubakov:2010qi} is instead closer to a single quark flavor in
a chirally broken phase where right and left chiralities are
living on the two boundaries of a single brane.}  
It consists of taking two gauge fields corresponding to the two chiralities
for each quark flavor
in a five dimensional AdS black hole background. 

The action is given by two Maxwell actions for left and right gauge fields 
plus separate Chern Simons terms corresponding to separate (``consistent'') anomalies
for left and right chiral quarks. The Chern-Simons terms are, however, not unique but
can be modified by adding total derivatives. A total derivative which
enforces invariance under vector gauge transformations $\delta V_M = \partial_M \lambda_V$
corresponds to
the so-called Bardeen counter-term \cite{Bardeen:1969md,Hill:2006ei}, 
leading to the action
\begin{eqnarray}\label{eq:action}
S &=& \int \sqrt{-g} \left( -\frac{1}{4 g_V^2} F^V_{MN} F_V^{MN}-\frac {1}{4 g_A^2} F^A_{MN} F_A^{MN}  \right.\nonumber\\&& 
 \left.+\frac{\kappa}{2}
\epsilon^{MNPQR} A_M ( F^A_{NP} F^A_{QR} + 3 F^V_{NP} F^V_{QR} ) \right).
\end{eqnarray}
Since the Chern-Simons term depends  explicitly on the gauge potential $A_M$, the action
is gauge invariant under $\delta A_M = \partial_M \lambda_A$ only up to a boundary term. This non-invariance is the holographic
implementation of the axial $U(1)$ anomaly in covariant (Adler-Bell-Jackiw) form
\cite{Adler:1969gk,Bell:1969ts} when identifying the gauge fields as holographic
sources for the currents of global $U(1)$ symmetries in the dual field theory. 
A rigorous string-theoretical realization of such a setup is provided for example
by the Sakai-Sugimoto model \cite{Sakai:2004cn,Sakai:2005yt}.
As usually done in the latter, we neglect the back-reaction of the bulk gauge fields
on the black hole geometry.

Before we proceed, we also want to clarify our conventions concerning the $\epsilon$-tensor. We define the $\epsilon$ tensor
$\epsilon_{MNPQR} = \sqrt{-g} \epsilon(MNPQR)$. Here we distinguish between
the tensor and the symbol. The symbol is $\epsilon(MNPQR)$ and normalized to $\epsilon(r0123)=1$.

In order to compute the field equations and the boundary action,
from which we shall obtain the two- and three-point functions of various currents,
we expand around fixed background gauge fields to second order in fluctuations.
The gauge fields are written as
\begin{eqnarray}
A_M = \cA_M + a_M\,,\qquad V_M = \cV_M + v_M,
\end{eqnarray}
where the calligraphic letters are the background fields and the lower case letters are the fluctuations.

After a little algebra, we find to first order in the fluctuations
\begin{eqnarray}\label{eq:firstorderaction}
\delta S^{(1)}_{\mathrm{bulk}+\partial} &=&  \int dr d^4x \sqrt{-g} \left\lbrace  a_M\left[ \frac{1}{g_A^2}\nabla_N \cF_A^{NM} + \frac{3\kappa}{2} \epsilon^{MNPQR}(  \cF_{NP}^A 
\cF_{QR}^A + \cF_{NP}^V \cF_{NP}^V)\right] \right. \nonumber \\
&& \quad + v_M\left.\left[ \frac{1}{g_V^2}\nabla_N \cF_V^{NM} + 3\kappa \epsilon^{MNPQR}(\cF_{NP}^A 
\cF_{QR}^V)\right] \right\rbrace\\
&+&  \int_{\partial} d^4x \left[a_\mu \left( \frac{1}{g_A^2}\sqrt{-g}\cF_A^{\mu r} +2 \kappa  \epsilon^{\mu\nu\rho\lambda} A_\nu \cF^A_{\rho\lambda}\right)
 \right.\nonumber \\
&& \quad +\left. v_\mu\left( \frac{1}{g_V^2}\sqrt{-g}\cF_V^{\mu r} + 6 \kappa \epsilon^{\mu\nu\rho\lambda} A_\nu \cF^V_{\rho\lambda}\right)\right]\nonumber
\,.
\end{eqnarray}

From the bulk term we get the equations of motion and from the boundary terms we can read the expressions for the currents,
\begin{eqnarray}\label{eq:currents}
J^\mu &=& \lim_{r\rightarrow\infty} \frac{1}{g_V^2}\sqrt{-g}\cF_V^{\mu r} + 6 \kappa \epsilon^{\mu\nu\rho\lambda} A_\nu \cF^V_{\rho\lambda}\,,\\
J_5^\mu &=&  \lim_{r\rightarrow\infty} \frac{1}{g_A^2}\sqrt{-g}\cF_A^{\mu r} +2 \kappa  \epsilon^{\mu\nu\rho\lambda} A_\nu \cF^A_{\rho\lambda}\,.
\end{eqnarray}
On-shell {they} obey
\begin{eqnarray}\label{eq:anomalies}
\partial_\mu J^\mu &=& 0\,,\nonumber\\
\partial_\mu J_5^\mu &=& - \frac{\kappa}2 \epsilon^{\mu\nu\rho\lambda} \left(3 F_{\mu\nu}^V F_{\rho\lambda}^V + F_{\mu\nu}^A F_{\rho\lambda}^A\right)\,.
\end{eqnarray}
As expected, the vector like current is exactly conserved. Comparing with the standard result from the one loop triangle calculation, we find
$\kappa = -\frac{N_c}{24 \pi^2}$ for a dual 
strongly coupled $SU(N_c)$ gauge theory for a 
massless Dirac fermion in the fundamental representation.

We emphasize that only by demanding an exact conservation law for the
vector current can we consistently couple it to an (external) electromagnetic field.
This leaves no ambiguity in the definitions of the above currents as the
ones obtained by varying the action with respect to the gauge fields and which
obey (\ref{eq:anomalies}). In particular,
we have to keep the contributions from the Chern-Simons terms in the action,
which are occasionally ignored in holographic calculations.

The second order term in the expansion of the action is
\begin{eqnarray}
S^{(2)}_{\mathrm{bulk}+\partial} &=& \int_{\mathrm{bulk}} \left\lbrace  a_M\left[ \frac{1}{2g_A^2} \nabla_N f_A^{NM} + \frac{3\kappa }{2} \epsilon^{MNPQR}(\cF_{NP}^A f_{QR}^A +  f_{NP}^V \cF_{QR}^V )\right] \right. \nonumber \\
&& \quad +   v_M\left.\left[ \frac{1}{2g_V^2} \nabla_N f_V^{NM} + \frac{3\kappa}{2} \epsilon^{MNPQR}(f_{NP}^A \cF_{QR}^V + f_{NP}^V \cF_{QR}^A)\right] \right\rbrace\\
&+& \int_{\partial}  \left[\frac{\sqrt{-g}}{2}(\frac{1}{g_A^2}a_\mu f_A^{\mu r} + \frac{1}{g_V^2}v_\mu f_V^{\mu r}) +  \kappa \epsilon^{\mu\nu\rho\lambda} 
( \cA_\nu a_\mu f^A_{\rho\lambda} + 3 v_\mu \cA_\nu f^V_{\rho\lambda} + 3 v_\mu a_\nu \cF^V_{\rho\lambda} ) \right]\,,\nonumber
 \end{eqnarray}
where $f_{MN}$ is the field strength of the fluctuations. Again, the action is already in the form of bulk equations of motion plus boundary term.

As gravitational background, we take the planar AdS Schwarzschild metric 
\begin{equation}
ds^2 = -f(r) dt^2 + \frac{dr^2}{f(r)} + \frac{r^2}{L^2} (dx^2+dy^2+dz^2)\, ,
\end{equation}
with $f= \frac{r^2}{L^2}-\frac{r_H^4}{r^2}$. The temperature is given in terms of the horizon by $r_H = L^2\pi T$. We rescale the $r$ coordinate such that the horizon lies at $r=1$ and we also will set the AdS scale $L=1$. Furthermore, we also rescale time and space coordinates accordingly. To recover the physical values of frequency and momentum we thus have to do replace 
$(\omega, k) \rightarrow (\omega/(\pi T) , k/(\pi T) )$.

The background gauge fields are
\begin{eqnarray}\label{eq:gaugebg}
\cA_0(r) &=& \Phi(r) = \alpha-\frac{\beta}{r^2}\,,\\
\label{eq:gaugebg2}
\cV_0(r) &=& \Psi(r) = \nu-\frac{\gamma}{r^2}\,.
\end{eqnarray}
We need to relate the integration constants $\alpha, \beta, \gamma,\nu$ to physical
observables now\footnote{Note that these integration constants with respect to the radial integration are independent of $(t,x,y,z)$!}. It is often stated in the literature that one needs to choose a gauge in which the fields in Eqs.~(\ref{eq:gaugebg}) and (\ref{eq:gaugebg2})
vanish on the horizon in order to make $A_M A^M$ and $V_M V^M$ well defined there. This is, however, not a physical constraint.
After all, the value of a gauge field has no intrinsic meaning. 

We will, instead, define
the chemical potentials of the global $U(1)$ symmetries 
as the potential difference between the horizon and
the boundary \cite{Ghoroku:2007re}.
This can be expressed as the integrated radial electric flux between horizon and boundary and is therefore a manifestly gauge invariant
quantity,
\begin{equation}\label{eq:defmu}
\mu   =  \int_{r_H=1}^{r_B=\infty}\partial_r A_0 dr = A_0(B) - A_0(H) \,,
\end{equation}
where $A_0$ stands for a generic gauge potential. The variation of the chemical potential can be thought of as either being a variation of the gauge potential on the boundary, the horizon
or an arbitrary combination thereof. However, by a gauge transformation we can always think of $\delta\mu$ to result from a variation
that vanishes on the horizon. Then $\delta \mu$ is just a special case of the general gauge field variation (\ref{eq:firstorderaction}), if
we interpret $a_\mu$, $v_\mu$ to be variations of the background fields. We see, therefore, that this definition automatically reproduces
$\frac{\delta S}{\delta \mu} = \langle Q \rangle$ where $Q$ is the integrated charge density $J^0$. In general, a variation of $\mu$ is
different from a variation with respect to the vector field. A variation in $\mu$ changes the ground state, $\delta \mu: ~ |Q\rangle \rightarrow
|Q+\delta Q\rangle$, whereas $\delta/\delta A_0$ inserts the operator $J^0$ into correlation functions. 

We can think of (\ref{eq:defmu}) as the difference of energy in the system with a unit of charge at the boundary and a unit
of charge at the horizon. This is the cost of energy to add a unit charge to the system and by definition represents thus
the chemical potential. {By the definition (\ref{eq:defmu}),} the integration constants $\beta$ and $\gamma$ are thus fixed to 
\begin{eqnarray}\label{eq:chempot}
\beta &=&  \mu_5,\\
\gamma &=&  \mu,
\end{eqnarray}
where $\mu$ is the chemical potential of the vector symmetry and $\mu_5$ the chemical potential of the axial $U(1)$. 
The constants $\alpha$ and $\nu$ we take {to be} arbitrary {and} we will eventually consider them as sources for insertions of the operators
$J^0$ and $J_5^0$ at zero momentum. Due to our choice of coordinates, the physical value of the chemical potentials is 
recovered by $\mu \rightarrow \pi T \mu$.

We can now compute the charges present in the system {from
the zero components of the currents   (\ref{eq:currents}) },
\begin{eqnarray}
J^0 &=& \frac{2\gamma}{g_V^2}, \\
J^0_5 &=& \frac{2\beta}{g_A^2}.
\end{eqnarray}
{Note that this is, in fact, the standard holographic definition {in} the grand canonical ensemble. Often the gauge choice
$A_0(H)=0$ is imposed from the outset and that fixes the integration constants $\alpha$ and $\nu$ to take the values
of the chemical potentials. It is important to realize that without a Chern-Simons term, the action for a gauge field in the bulk 
depends only on the field strengths and is, therefore, independent of constant boundary values of the gauge field. The
action does, of course, depend on the physically 
{relevant and gauge invariant}
difference of the potential between the horizon and the boundary. 
For our particular model, the choice of {the} Chern-Simons term results, however, also in an explicit dependence on the integration constant $\alpha$. It is crucial to keep in mind that $\alpha$ is a priori unrelated to the chiral chemical potential but plays the
role of the source for the operator $J_5^0$ at zero momentum.}

For the fluctuations we choose the gauge $a_r=0$. We take the fluctuations to be of plane wave form with frequency $\omega$ and
momentum $k$ in $x$-direction. The relevant polarizations are then the $y$- and $z$-components, i.e. the transverse gauge field 
fluctuations.
The equations of motion are
\begin{eqnarray}
v_i'' + (\frac{f'}{f}+\frac{1}{r}) v_i' +\frac{(\omega^2r^2-fk^2)}{f^2r^2} v_i + \frac{12i \kappa g_V^2 k}{f r} \epsilon_{ij} ( \Phi' v_j + \Psi' a_j) &=& 0\,, \\
a_i'' + (\frac{f'}{f}+\frac{1}{r}) a_i' +\frac{(\omega^2r^2-fk^2)}{f^2r^2} a_i + \frac{12i \kappa g_A^2 k}{f r} \epsilon_{ij} ( \Phi' a_j + \Psi' v_j) &=& 0\,.
\end{eqnarray}
The indices $(i,j) \in \{y,z\}$ and a prime denotes differentiation with respect to the radial coordinate $r$. The two-dimensional
epsilon symbol is $\epsilon_{yz} =1$.

There is also a longitudinal sector of gauge field equations. They receive no contribution from the Chern-Simons term and so are uninteresting for our purposes.

The boundary action in Fourier space in the relevant transversal sector is
\begin{eqnarray}
S^{(2)} = \int_{\partial} dk \left[ -\frac{r f}{2}  (\frac{1}{g_A^2}a_{-k}^i (a_k^i)' + \frac{1}{g_V^2}v_{-k}^i (v_k^i)')-
2 i k  \kappa\epsilon_{ij}\alpha \left( a^i_{-k} a^j_{k}  + 3 v^i_{-k} v^j_{k}\right)\right]
\end{eqnarray}
As anticipated, the second order boundary action depends on the boundary value of the axial gauge field but not on the boundary value of the vector gauge field.

From this we can compute the holographic Green function. The way to do this is to compute four linearly independent solutions that
satisfy in-falling boundary conditions on the horizon \cite{Son:2002sd,Herzog:2002pc}. At the AdS boundary we require that the first solution asymptotes to the vector $(v_y, v_z, a_y, a_z) = 
(1,0,0,0)$, the second solution to the vector $(0,1,0,0)$ and so on. We can therefore build up a matrix of solutions $F_k\,^I\,_J(r)$ where each column corresponds to one of these solutions \cite{Kaminski:2009dh}. Given a set of boundary fields $a_i^{(0)}(k)$, $v_i^{(0)}(k)$, which we collectively arrange in the vector $\varphi^{I,(0)}(k)$, the bulk solution corresponding to these boundary fields is 
\begin{equation}
\varphi^I(k,r) = F_k^I\,_J \varphi^{J,(0)}(k).
\end{equation}
Here, $F$ is the (matrix valued) bulk-to-boundary propagator for the system of coupled differential equations. 

The holographic Green function is then given by
\begin{eqnarray}
G_{IJ} =  - 2\lim_{r\rightarrow \infty} (\bbA_{IL} (F_k\,^L\,_J)' + \bbB_{IJ} ).
\end{eqnarray}
The matrices $\bbA$ and $\bbB$
 can be read off from the boundary action as
\begin{equation}
\bbA = -\frac 1 2 r f  \left( 
\begin{array}{cc}
\frac{1}{g_V^2} &  0\\ 
0& \frac{1}{g_A^2}\\ 
\end{array}  
\right)~~~,~~~~~\bbB = -2 i\kappa k \alpha  \left( 
\begin{array}{cc}
3 \epsilon_{ij} &  0\\ 
0&  \epsilon_{ij}\\ 
\end{array}  
\right),
\end{equation}
(notice that $F$ becomes the unit matrix at the boundary).

We are interested here only in the zero frequency limit and the first order in an expansion in the momentum $k$. In this limit, the
differential equations can be solved explicitly. 
To this order the matrix bulk-to-boundary propagator is
\begin{equation}
F = \left(
\begin{array}{cccc}
1 &  -g_A^2\mu_5 g(r) & 0 &  -g_V^2\mu g(r) \\
 g_A^2\mu_5 g(r) & 1&   g_V^2\mu g(r)&0 \\
0 & -g_V^2\mu g(r) & 1 &  -g_A^2\mu_5 g(r) \\
 g_V^2\mu g(r) & 0 &  g_A^2\mu_5 g(r)&1 \\
\end{array}
\right),
\end{equation}
where $g(r) =  6 i k \kappa \log(1+1/r^2) $.
We find then the holographic current two-point functions
\begin{eqnarray}\label{eq:correlators}
\langle J^i J^j \rangle &=& -12 i \kappa k (\mu_5 - \alpha) \epsilon_{ij},\\
\langle J_5^i J^j \rangle &=& -12 i \kappa k \mu \epsilon_{ij},\\
\label{eq:correlators2}
\langle J_5^i J^j_5 \rangle &=& -4 i \kappa k (3\mu_5 - \alpha) \epsilon_{ij}.
\end{eqnarray}
Although $\mu,\mu_5$ and the boundary gauge field value $\alpha$ enter in very similar ways in this result, we need to remember
their completely different physical meaning. The chemical potentials $\mu$ and $\mu_5$ are gauge invariant physical state variables whereas
$\alpha$ is {the source for insertions of $J_5^0(0)$}. Had we chosen the ``gauge'' $\alpha=\mu_5$, we would have concluded
(erroneously) that the two-point correlator of electric currents vanishes. We see now that with $\mu_5$ introduced separately from $\alpha$ {that} this is not so.
We simply have obtained expressions for the correlators in the physical state described by $\mu$ and $\mu_5$ in the constant
external background field $\alpha$. Due to the gauge invariance of the action under vector gauge transformations,
the constant mode $\nu$ of the corresponding source does not appear. The physical difference between the chemical potentials and the gauge field values is
clear now. The susceptibilities of the two-point functions obtained by differentiating with respect to the
chemical potentials are different from the three-point functions obtained by differentiating with respect to the gauge field
values. Finally, we remark that the temperature dependence drops out due to the opposite scaling of $k$ and $\mu$, $\mu_5$.

To compute the anomalous conductivities we therefore have to evaluate the two-point function 
for vanishing background fields $\nu=\alpha=0$. We obtain, in complete agreement with the well-known weak coupling results,
\begin{eqnarray}\label{eq:conductivities}
J^i=e^2 \sigma_{\rm CME} B^i,\qquad
\sigma_{\rm CME} &=& \lim_{k\to0}\frac{i\epsilon_{ij}}{2k}\langle J^i J^j \rangle|_{\nu=\alpha=0}
=\frac{N_c 
}{2\pi^2} \mu_5, \\
J^i_5=e \sigma_{\mathrm{axial}} B^i,\qquad\;
\sigma_{\mathrm{axial}} &=& \lim_{k\to0}\frac{i\epsilon_{ij}}{2k}\langle J_5^i J^j \rangle|_{\nu=\alpha=0}
=\frac{N_c 
}{2\pi^2} \mu,\\ \label{eq:conductivities3}
J^i_5=\sigma_{55} B_5^i,\qquad\quad
\sigma_{55} &=& \lim_{k\to0}\frac{i\epsilon_{ij}}{2k}\langle J_5^i J_5^j \rangle|_{\nu=\alpha=0}
=\frac{N_c 
}{2\pi^2} \mu_5.
\end{eqnarray}
We are tempted to call all $\sigma$'s conductivities. This is, however, a slight misuse of language in the case of $\sigma_{55}$.
Formally, $\sigma_{55}$ measures the response due to the presence of an axial magnetic field $\vec{B}_5 = \nabla \times \vec{A_5}$. 
Since such fields do not exist in nature, we cannot measure $\sigma_{55}$ in the same way as $\sigma_{CME}$ and  $\sigma_{axial}$.

Since the two-point functions in Eqs.~(\ref{eq:correlators})-(\ref{eq:correlators2}) still depend on the external source $\alpha$, we can also obtain the three
point functions in a particular kinematic regime. Differentiating with respect to $\alpha$ (and $\nu$) we find the three
point functions

\begin{eqnarray}\label{eq:threepts1}
\langle J^i(k) J^j(-k) J^0(0)\rangle &=& 0, \\ \label{eq:threepts2}
\langle J_5^i(k) J^j(-k) J^0(0)\rangle &=& 0, \\ \label{eq:threepts3}
\langle J_5^i(k) J_5^j(-k) J^0(0)\rangle &=& 0, \\ \label{eq:threepts4}
\langle J^i(k) J^j(-k) J_5^0(0)\rangle &=& -i k \frac{N_c 
}{2\pi^2} \epsilon_{ij}, \label{eq:jjj5}\\ \label{eq:threepts5} 
\langle J_5^i(k) J^j(-k) J_5^0(0)\rangle &=& 0, \\ \label{eq:threepts6}
\langle J_5^i(k) J_5^j(-k) J_5^0(0)\rangle &=& -i k \frac 1 3  \frac{N_c 
}{2\pi^2} \epsilon_{ij}.
\end{eqnarray}
Note the independence on chemical potentials and temperature. 



{Equations (\ref{eq:threepts4}) and (\ref{eq:threepts6}) show the sensitivity of the theory to a constant temporal component of
the axial gauge field even at zero temperature and chemical potentials. If the axial} $U(1)$ {symmetry was exactly
conserved, such a constant field value would be a gauge degree of freedom and the theory would be insensitive to it. 
Since this symmetry is, however, anomalous, it couples {to currents} through these three-point functions.
The correlators (\ref{eq:threepts4}) and (\ref{eq:threepts6}) can therefore be understood as expressing the anomaly in the 
axial} $U(1)$ {symmetry.}

In the next section we will check these results in vacuum at weak coupling by calculating the triangle diagram in the
relevant kinematic regimes.

\section{Weak-coupling calculations}

An important property of the two- and three-point functions we just calculated is that they are independent of temperature.
The three-point functions are furthermore independent of the chemical potentials. Therefore, the results for the three-point
functions should coincide with correlation functions in (a chirally symmetric) vacuum. At weak coupling, all the three-point functions can be 
obtained from a single 1-loop Feynman integral. We only need to evaluate the diagram with two vector currents and
one axial current. The diagram with three vector currents vanishes identically {(due to C-parity)} and the one with three axial currents
can be reduced to the one with only one axial current by anti-commuting $\gamma_5$ matrices (when a regularization is applied that permits this). Similarly, it can be seen that the diagram with two axial vector currents can be reduced to the one with none, which vanishes.

When computing the three-point function, it is crucial to check the resulting anomalies. Gauge invariant regulators, like dimensional regularization, should yield the correct covariant anomaly, such that the vector currents are identically conserved. On the other hand, for example cutoff regularization breaks gauge invariance and further finite renormalizations may be needed in order to restore gauge invariance. In the following, we apply both dimensional and cutoff regularizations to compute the three-point function and show that they give consistent results with each other and with Eqs.~(\ref{eq:threepts1})-(\ref{eq:threepts6}).

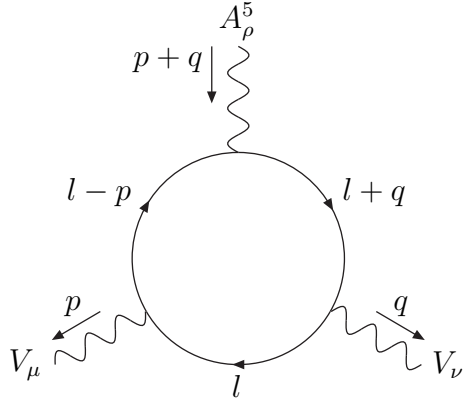
\begin{figure}
\centering
 \begin{picture}(300,140)(0,0)
  \Photon(81,10)(115,30){4}{3}
  \Photon(185,30)(219,10){4}{3}
  \Photon(150,130)(150,90){4}{3}
  \ArrowArcn(150,50)(40,90,-30)
  \ArrowArcn(150,50)(40,210,90)
  \ArrowArcn(150,50)(40,330,210)
  \LongArrow(98,30)(81,20)
  \LongArrow(202,30)(219,20)
  \LongArrow(140,130)(140,110)
  \Text(70,10)[]{$V_\mu$}
  \Text(230,10)[]{$V_\nu$}
  \Text(150,140)[]{$A^5_\rho$}
  \Text(88,32)[]{$p$}
  \Text(212,32)[]{$q$}
  \Text(137,130)[rt]{$p+q$}
  \Text(150,2)[]{$l$}
  \Text(110,70)[rb]{$l-p$}
  \Text(190,70)[lb]{$l+q$}
 \end{picture}
\caption{The triangle diagram.}
\label{fig:triangle}
\end{figure}

\subsection{Triangle diagram with one axial current}
 
The triangle diagram, shown in Fig.~\ref{fig:triangle}, with one axial current and two vector currents, is given by
\begin{eqnarray}\label{eq:triangle}
\Gamma^{\mu\nu\rho}(p,q) &=& (-1) (i e)^2 (i g) (i)^3 \int \frac{d^d l}{(2\pi)^d} \mathrm{tr}\left( 
\gamma_5 
 \frac{\slashed{l}-\slashed{p}}{(l-p)^2}
\gamma^\mu
\frac{ \slashed{l}}{l^2} \gamma^\nu \frac{\slashed{l}+\slashed{q}}{(l+q)^2} \gamma^\rho  \right) \nonumber\\
 & &+(\mu \leftrightarrow \nu , p \leftrightarrow q).
\end{eqnarray}
The factors are a $(-1)$ from the fermion loop, the couplings to vector and axial gauge fields and $i$ for
each fermion propagator. We will simply set the electric and axial couplings $e$ and $g$ to one. Evaluation of the integral with dimensional and cutoff regularizations is presented in some detail in Appendix \ref{app:triangle}.

The anomalies of the various currents coupled to the triangle diagram are obtained by contracting the three-point function above by the corresponding momenta. Applying dimensional regularization, we get immediately
\begin{eqnarray}
 p_\mu \Gamma^{\mu\nu\rho}_\mathrm{DR}(p,q) &=&0,\\
 q_\nu \Gamma^{\mu\nu\rho}_\mathrm{DR}(p,q) &=&0,\\
 (p+q)_\rho \Gamma^{\mu\nu\rho}_\mathrm{DR}(p,q)&=&  \frac{i}{2\pi^2} p_\alpha q_\beta \epsilon^{\alpha\beta\mu\nu},
 \end{eqnarray}
 yielding the correct Adler-Bell-Jackiw anomaly. In terms of cutoff regularization, we, however, find
 \begin{eqnarray}
 p_\mu \Gamma^{\mu\nu\rho}_\mathrm{CO}(p,q) &=&-\frac{i}{6\pi^2}p_\alpha q_\beta\epsilon^{\alpha\beta\nu\rho},\\
 q_\nu \Gamma^{\mu\nu\rho}_\mathrm{CO}(p,q) &=&\frac{i}{6\pi^2}p_\alpha q_\beta\epsilon^{\alpha\beta\mu\rho},\\
 (p+q)_\rho \Gamma^{\mu\nu\rho}_\mathrm{CO}(p,q)&=&  \frac{i}{6\pi^2} p_\alpha q_\beta \epsilon^{\alpha\beta\mu\nu}.
 \end{eqnarray}
 In order to cancel the anomalies in the vector current, we must perform an additional finite renormalization by adding the Bardeen counter-term,
 \begin{equation}
 \Gamma^{\mathrm{c.t.}} = c \int\, d^4x\, \epsilon^{\mu\nu\rho\lambda} \,V_\mu\, \,A^5_\nu \,F^V_{\rho\lambda},
 \end{equation}
 where $F^V_{\rho\lambda} = \partial_\rho V_\lambda -  \partial_\lambda  V_\rho$. This vertex brings an additional contribution to the three-point function, and the full result reads
\begin{equation}
\Gamma^{\mu\nu\rho} =  \Gamma_{\mathrm{CO}}^{\mu\nu\rho}(p,q) + 2 i c (p_\lambda - q_\lambda ) \epsilon^{\lambda\mu\nu\rho}.
\end{equation}
Choosing the coefficient $c$ of the Bardeen counter-term appropriately, $c=\frac{1}{12 \pi^2}$, we find the anomaly equations
\begin{eqnarray}
 p_\mu \Gamma^{\mu\nu\rho}(p,q) &=&0,\\
 q_\nu \Gamma^{\mu\nu\rho}(p,q) &=&0,\\
 (p+q)_\rho \Gamma^{\mu\nu\rho}(p,q)&=& \frac{i}{2\pi^2} p_\alpha q_\beta \epsilon^{\alpha\beta\mu\nu},
 \end{eqnarray}
in full agreement with the covariant anomaly and the result from dimensional regularization.

 We next want to evaluate the triangle diagram in the special kinematic regimes 
 of Eqs.~(\ref{eq:threepts1})-(\ref{eq:threepts6}). Taking $q=-p$, corresponding to the three-point function in Eq.~(\ref{eq:threepts4}), only the integrands $A$ and $B$ in Eqs.~(\ref{eq:coeffs_begin})-(\ref{eq:coeffs_end}) contribute  
 and take the values $1/2$ and $-1/2$ in dimensional regularization and $1/6$ and $-1/6$ in cutoff, respectively. The three-point function is then 
\begin{equation}
\Gamma^{\mu\nu 0}(p,-p)  = \frac{i}{2\pi^2} \epsilon^{\alpha\mu\nu 0} p_\alpha \,,
\end{equation}
in agreement with Eq.~(\ref{eq:threepts4}). Note that with cutoff regularization, $\frac 1 3$ of this result comes from the loop diagram and $\frac 2 3$ comes from the counter-term.

Let us next take $p=0$, {\it i.e.} we put zero momentum on one of the vector currents. The corresponding loop integral vanishes in dimensional regularization, while the loop contribution in cutoff regularization is precisely canceled by the contribution from the counter-term,
\begin{equation}
\Gamma^{0\nu\rho}(0,-q)=0\,.
\end{equation}
This result is in agreement with Eq.~(\ref{eq:threepts2})

\subsection{Triangle diagram with three axial currents}

From the same one loop integral we can also compute the correlator of three axial currents\footnote{However, as this requires commuting the $\gamma_5$ with the rest of the $\gamma$ matrices, only cutoff regularization can be applied.}. Since we can anti-commute the $\gamma_5$
and use $\gamma_5^2=-1$, we can reduce this diagram to the one in Fig.~(\ref{eq:triangle}). The Bardeen counter-term, however, does not contribute this time, and we therefore find
\begin{equation}
\Gamma_5^{\mu\nu 0} = \frac{1}{3} \frac{i}{2 \pi^2} \epsilon^{\alpha \mu\nu 0} p_\alpha  \,,
\end{equation}
just as in Eq.~(\ref{eq:threepts6}). The factor $\frac 1 3$ is fixed by demanding Bose symmetry on the external legs. 

All other current three-point functions can be related to the triangle with three vector currents which is known to vanish.
Therefore, we have indeed reproduced the holographic results in Eqs.~(\ref{eq:threepts1})-(\ref{eq:threepts6}).

\subsection{Two-point correlator of axial currents}

To conclude the weak coupling considerations, we also compute the two-point function of two axial currents in the background
of an axial chemical potential. We simply can follow the analogous calculation of the chiral magnetic conductivity in
\cite{Kharzeev:2009pj}. The relevant two-point function of axial currents is
\begin{equation}
G_{55}(P) = \frac{T}{2} \sum_{\tilde \omega_n} \int \frac{d^3 q}{(2\pi)^3} \epsilon^{ijk} 
\tr\left[ \gamma^k \gamma_5 S(Q) \gamma^j \gamma_5 S(P+Q)\right],
\end{equation}
where $S(Q)$ is the fermion propagator at finite temperature and density,
\begin{equation}
S(Q) = \frac{1}{i\gamma^0(\tilde \omega_n - i \mu -i \mu_5\gamma_5) - \gamma \vec{q}},
\end{equation}
with $\omega_n = (2n+1)\pi T$. Using $\gamma_5 S(Q) = - S(Q) \gamma_5$ we can square the $\gamma_5$ matrices to one and end up with the
same integral as for the chiral magnetic conductivity in \cite{Kharzeev:2009pj}.
We therefore find at weak coupling that $\sigma_{55} = \sigma_{CME}$, coinciding with the result in (\ref{eq:conductivities3}).

\section{Discussion and Conclusion}

We have computed two- and three-point functions of currents at finite density
using holographic methods for a simple holographic model incorporating the
axial anomaly of the standard model. We were able to reproduce the known
weak-coupling results {concerning the chiral magnetic effect} and also found a new type of
``conductivity'' in the axial sector alone, $\sigma_{55}$. Although it can
not be probed by switching on external fields,
as a two-point function it is as well defined as $\sigma_{CME}$.
It would be interesting to find a way of relating this anomalous
conductivity
to experimentally accessible observables.

Previous calculations of anomalous conductivities have been able
to reproduce the weak-coupling result for $\sigma_{\rm axial}\propto\mu$
but not $\sigma_{CME}\propto \mu_5$ unless the contributions from
the Chern-Simons term to the chiral currents were dropped.
In our calculation we have used the complete expressions for the
currents, but
of key importance was a clear distinction between the physical state
variable, the chemical potential, and the external background
field. The latter we viewed exclusively as a source that couples to an
operator, whereas the chemical potential should correspond in the most
elementary way to the cost of energy for adding a unit of charge to the
system.

It is useful to remember how a chemical potential can be introduced 
in field theory. One possible way is by deforming the Hamiltonian according
to $H\rightarrow H-\mu Q$. A second, usually equivalent, way is by
imposing boundary conditions $\phi(t - i\beta) = \pm \exp( \mu \beta) \phi(t)$
 on the fields along the imaginary time direction \cite{Landsman:1986uw,Evans:1995yz}.
These methods are equivalent as long as $Q$ is a non-anomalous charge.
Similarly, in holography we can introduce the chemical potential either through a boundary
value of the temporal component of the gauge field or through the potential difference
between boundary and horizon.
Thus, for non-anomalous symmetries, the boundary value of the temporal gauge field
can be identified with
the chemical potential. Due to the exact gauge invariance of the action, a
constant boundary value
never enters in correlation functions.
In the presence of a Chern-Simons term, however, the gauge symmetry is 
partially lost
and even a constant boundary gauge field becomes observable. This can be
seen explicitly
from the three-point functions (\ref{eq:threepts4}) and (\ref{eq:threepts6}).
Therefore, we should set the axial vector field to zero after
having used it as a source for axial current. By defining the corresponding
chemical potential as the potential difference between
the horizon of the AdS black hole and the holographic boundary we are able
to
do so. However, the prize we have to pay is to accept singular gauge field
configurations at the horizon.

The fact that a gauge field that does not vanish on the horizon is not
well defined is most easily seen in Kruskal coordinates,
\begin{equation}
UV = - \exp(4\pi T r_*) ~~~~,~~~~ V/U = -\exp(4 \pi T t)\,,
\end{equation}
where $dr_* = dr/f$. We note that close to the horizon, $r-r_H \approx -UV$.
The time component of a gauge field in Kruskal
coordinates at the horizon is therefore
\begin{equation}\label{eq:singular}
A_0 dt = A_0(r_H) ( \frac{dV}{V} - \frac{dU}{U}) - A_0'(r_H)( UdV-VdU) +
\cdots
\end{equation}
This is not a well-defined one-form unless $A_0(r_H)$ vanishes.
Although the gauge field is singular at the horizon, we do not believe that
well defined physical observables are effected by this. Local gauge invariant 
observables, i.e. the field strengths, are certainly well behaved.

\begin{acknowledgments}
We would like to thank Kenji Fukushima, Dima Kharzeev,
Valery Rubakov, Andreas Schmitt, Dam Son, Misha Stephanov,
Harmen Warringa, Larry Yaffe, Ho-Ung Yee, and Ariel Zhitnitsky for useful discussions.

This work has been supported by Accion Integrada Hispanoaustriaca HA2008-0003, Plan Nacional de Alta Energ\'\i as FPA-2006-05485, 
Comunidad de Madrid HEP-HACOS S2009/ESP-1473
and \"OAD, project no. ES12/2009.
AG acknowledges financial support 
by the Austrian Science Foundation FWF, project no.\ P22114
and by the Academy of Finland, project no.\ 136997.

\end{acknowledgments}

\appendix

\section{Evaluation of the triangle diagram}
\label{app:triangle}

We wish to compute the integral corresponding to the triangle diagram in Fig.~\ref{fig:triangle},
\begin{eqnarray}
\Gamma^{\mu\nu\rho}(p,q) &=& (-1) (i e)^2 (i g) (i)^3 \int \frac{d^d l}{(2\pi)^d} \mathrm{tr}\left( 
\gamma_5 
 \frac{\slashed{l}-\slashed{p}}{(l-p)^2}
\gamma^\mu
\frac{ \slashed{l}}{l^2} \gamma^\nu \frac{\slashed{l}+\slashed{q}}{(l+q)^2} \gamma^\rho  \right) \nonumber\\
 & &+(\mu \leftrightarrow \nu , p \leftrightarrow q).
\end{eqnarray}
Using Feynman parametrization the integral can be written as
\begin{eqnarray}
\Gamma^{\mu\nu\rho}(p,q) &=&  I_{\alpha \beta \gamma} \left[ \mathrm{tr} \left(
\gamma_5\gamma^\alpha \gamma^\mu \gamma^\beta \gamma^\nu \gamma^\gamma \gamma^\rho \right)
- \mathrm{tr} \left(
\gamma_5\gamma^\gamma \gamma^\nu \gamma^\beta \gamma^\mu \gamma^\alpha \gamma^\rho \right) \right], \\
I_{\alpha\beta\gamma} &=& -2 \int_0^1 dx dy\, \Theta(1-x-y) \int  \frac{d^d l}{(2\pi)^d}  \frac{N_{\alpha\beta\gamma}}{(l^2+D)^3},
\end{eqnarray}
where
\begin{eqnarray}
D &=& x(1-x) p^2 + 2 x y p\cdot q + y(1-y) q^2, \\
r_\mu &=& x p_\mu - y q_\mu, \\
N_{\alpha\beta\gamma} &=& (r-p)_\alpha r_\beta (r+q)_\gamma + \frac{l^2}{d}\Big[ \delta_{\alpha\beta}(r+q)_\gamma +   \delta_{\alpha\gamma}\,r_\beta + \delta_{\beta\gamma}(r-p)_\alpha \Big].
\end{eqnarray}
Here we have already taken into account that with both dimensional and cutoff regularizations, the integral with odd powers of $l$ in the numerator of the integrand vanishes, and the remaining tensor structure is dictated by the rotational symmetry of a momentum shell at fixed $|l|$.

Using
\begin{eqnarray}
\int_0^\Lambda \frac{l^3 dl}{(l^2+D)^3} &=& \frac{1}{4D} + \mathcal{O}\left(\frac{1}{\Lambda^2}\right), \\
\int_0^\Lambda \frac{l^5 dl}{(l^2+D)^3} &=&
\frac 1 2 \left[ \log\left(\frac{\Lambda^2}{D}\right) - \frac 3 2 \right]  + \mathcal{O}\left(\frac{1}{\Lambda^2}\right),
\end{eqnarray}
in the cutoff regularization ($d=4$), and
\begin{eqnarray} 
\left(\frac{\mathrm{e}^{\gamma_E} \mubar^2}{4\pi}\right)^\epsilon\,\int\,\frac{d^{4-2\epsilon}l}{(2\pi)^{4-2\epsilon}}\frac{1}{(l^2+D)^3} & = & \frac{\Gamma(\epsilon)}{16\pi^2}\left(\mathrm{e}^{\gamma_E} \mubar^2 \right)^\epsilon \frac{\epsilon}{2}\frac{1}{D^{1+\epsilon}}, \\
\left(\frac{\mathrm{e}^{\gamma_E} \mubar^2}{4\pi}\right)^\epsilon\,\int\,\frac{d^{4-2\epsilon}l}{(2\pi)^{4-2\epsilon}}\frac{l^2}{(l^2+D)^3} & = & \frac{\Gamma(\epsilon)}{16\pi^2}\left(\mathrm{e}^{\gamma_E} \mubar^2 \right)^\epsilon \left(1-\frac{\epsilon}{2}\right)\frac{1}{D^\epsilon},
\end{eqnarray}
in the dimensional regularization ($d=4-2\epsilon$) with $\ms$ scheme, we find
\begin{eqnarray}
\Gamma^{\mu\nu\rho}_\mathrm{reg}(p,q) &=& \frac{i}{2 \pi^2} \int_0^1\, dx\, dz\,\Theta(1-x-z) \Big[\,
(A^\mathrm{reg} p_\alpha + B^\mathrm{reg} q_\alpha) \epsilon^{\alpha \mu\nu\rho} \nonumber \\
& & +
 ( C_1^\mathrm{reg} p^\mu + D_1^\mathrm{reg} q^\mu) p_\alpha q_\beta \epsilon^{\alpha\beta\nu\rho} +
( C_2^\mathrm{reg} p^\nu + D_2^\mathrm{reg} q^\nu) p_\alpha q_\beta \epsilon^{\alpha\beta\mu\rho} \Big],
\end{eqnarray}
with $\mathrm{reg}\in\{\mathrm{CO},\mathrm{DR}\}$. The coefficients are given by
\begin{eqnarray}\label{eq:coeffs_begin}
A^\mathrm{CO} &=&  \frac{ (x-1)r^2 + y q^2}{D} + \left[ \log\left(\frac{\Lambda^2}{D}\right) - \frac 3 2 \right] (3x-1),\\
B^\mathrm{CO} &=& \frac{ (1-y)r^2 - x p^2}{D} + \left[ \log\left(\frac{\Lambda^2}{D}\right) - \frac 3 2 \right] (1-3y), \\
C_1^\mathrm{CO} & =& \frac{2 x(x-1)}{D},  \\
C_2^\mathrm{CO} &=& \frac{2 x y}{D},  \\
D_1^\mathrm{CO} &=& -\frac{2 x y}{D}, \\
D_2^\mathrm{CO} &=& \frac{2 y (1-y)}{D},
 \end{eqnarray}
in the cutoff regularization, and
\begin{eqnarray}
A^\mathrm{DR} & = & \left[\frac{(x-1)(r^2-D) + yq^2}{D}\epsilon+(3x-1) \right]\frac{\Gamma(\epsilon)}{D^{\epsilon}}\left(\mathrm{e}^{\gamma_E} \mubar^2 \right)^\epsilon, \\
B^\mathrm{DR} & = & \left[\frac{(1-y)(r^2-D) - xp^2}{D}\epsilon + (1-3y) \right]\frac{\Gamma(\epsilon)}{D^{\epsilon}}\left(\mathrm{e}^{\gamma_E} \mubar^2 \right)^\epsilon, \\
C_1^\mathrm{DR} & = & \frac{2\epsilon x(x-1)}{D^{1+\epsilon}} \Gamma(\epsilon)\left(\mathrm{e}^{\gamma_E} \mubar^2 \right)^\epsilon, \\
C_2^\mathrm{DR} & = & \frac{2\epsilon xy}{D^{1+\epsilon}}\Gamma(\epsilon)\left(\mathrm{e}^{\gamma_E} \mubar^2 \right)^\epsilon, \\
D_1^\mathrm{DR} & = & -\frac{2\epsilon xy}{D^{1+\epsilon}}\Gamma(\epsilon)\left(\mathrm{e}^{\gamma_E} \mubar^2 \right)^\epsilon, \\ \label{eq:coeffs_end}
D_2^\mathrm{DR} & = & -\frac{2\epsilon y(y-1)}{D^{1+\epsilon}} \Gamma(\epsilon)\left(\mathrm{e}^{\gamma_E} \mubar^2 \right)^\epsilon,
\end{eqnarray}
in the dimensional regularization.

\bibliographystyle{JHEP}
\bibliography{cmch}

\providecommand{\href}[2]{#2}\begingroup\raggedright\begin{thebibliography}{10}

\bibitem{Son:2004tq}
D.~T. Son and A.~R. Zhitnitsky, {\it {Quantum anomalies in dense matter}},
  {\em Phys. Rev.} {\bf D70} (2004) 074018,
  [\href{http://xxx.lanl.gov/abs/hep-ph/0405216}{{\tt hep-ph/0405216}}].

\bibitem{Metlitski:2005pr}
M.~A. Metlitski and A.~R. Zhitnitsky, {\it {Anomalous axion interactions and
  topological currents in dense matter}},  {\em Phys. Rev.} {\bf D72} (2005)
  045011, [\href{http://xxx.lanl.gov/abs/hep-ph/0505072}{{\tt
  hep-ph/0505072}}].

\bibitem{Newman:2005as}
G.~M. Newman and D.~T. Son, {\it {Response of strongly-interacting matter to
  magnetic field: Some exact results}},  {\em Phys. Rev.} {\bf D73} (2006)
  045006, [\href{http://xxx.lanl.gov/abs/hep-ph/0510049}{{\tt
  hep-ph/0510049}}].

\bibitem{Charbonneau:2009ax}
J.~Charbonneau and A.~Zhitnitsky, {\it {Topological Currents in Neutron Stars:
  Kicks, Precession, Toroidal Fields, and Magnetic Helicity}},  {\em JCAP} {\bf
  1008} (2010) 010, [\href{http://xxx.lanl.gov/abs/0903.4450}{{\tt
  arXiv:0903.4450}}].

\bibitem{Kharzeev:2004ey}
D.~Kharzeev, {\it {Parity violation in hot QCD: Why it can happen, and how to
  look for it}},  {\em Phys. Lett.} {\bf B633} (2006) 260--264,
  [\href{http://xxx.lanl.gov/abs/hep-ph/0406125}{{\tt hep-ph/0406125}}].

\bibitem{Kharzeev:2007tn}
D.~Kharzeev and A.~Zhitnitsky, {\it {Charge separation induced by P-odd bubbles
  in QCD matter}},  {\em Nucl. Phys.} {\bf A797} (2007) 67--79,
  [\href{http://xxx.lanl.gov/abs/0706.1026}{{\tt arXiv:0706.1026}}].

\bibitem{Kharzeev:2007jp}
D.~E. Kharzeev, L.~D. McLerran, and H.~J. Warringa, {\it {The effects of
  topological charge change in heavy ion collisions: 'Event by event P and CP
  violation'}},  {\em Nucl. Phys.} {\bf A803} (2008) 227--253,
  [\href{http://xxx.lanl.gov/abs/0711.0950}{{\tt arXiv:0711.0950}}].

\bibitem{Fukushima:2008xe}
K.~Fukushima, D.~E. Kharzeev, and H.~J. Warringa, {\it {The Chiral Magnetic
  Effect}},  {\em Phys. Rev.} {\bf D78} (2008) 074033,
  [\href{http://xxx.lanl.gov/abs/0808.3382}{{\tt arXiv:0808.3382}}].

\bibitem{Kharzeev:2009pj}
D.~E. Kharzeev and H.~J. Warringa, {\it {Chiral Magnetic conductivity}},  {\em
  Phys. Rev.} {\bf D80} (2009) 034028,
  [\href{http://xxx.lanl.gov/abs/0907.5007}{{\tt arXiv:0907.5007}}].

\bibitem{Alekseev:1998ds}
A.~Y. Alekseev, V.~V. Cheianov, and J.~Fr{\"o}hlich, {\it {Universality of
  transport properties in equilibrium, Goldstone theorem and chiral anomaly}},
  {\em Phys. Rev. Lett.} {\bf 81} (1998) 3503--3506,
  [\href{http://xxx.lanl.gov/abs/cond-mat/9803346}{{\tt cond-mat/9803346}}].

\bibitem{Abelev:2009uh}
{\bf STAR} Collaboration, B.~I. Abelev {\em et.~al.}, {\it {Azimuthal
  Charged-Particle Correlations and Possible Local Strong Parity Violation}},
  {\em Phys. Rev. Lett.} {\bf 103} (2009) 251601,
  [\href{http://xxx.lanl.gov/abs/0909.1739}{{\tt arXiv:0909.1739}}].

\bibitem{Voloshin:2008jx}
{\bf STAR} Collaboration, S.~A. Voloshin, {\it {Probe for the strong parity
  violation effects at RHIC with three particle correlations}},
  \href{http://xxx.lanl.gov/abs/0806.0029}{{\tt arXiv:0806.0029}}.

\bibitem{Wang:2009kd}
F.~Wang, {\it {Effects of Cluster Particle Correlations on Local Parity
  Violation Observables}},  {\em Phys.Rev.} {\bf C81} (2010) 064902,
  [\href{http://xxx.lanl.gov/abs/0911.1482}{{\tt arXiv:0911.1482}}].

\bibitem{Asakawa:2010bu}
M.~Asakawa, A.~Majumder, and B.~Muller, {\it {Electric Charge Separation in
  Strong Transient Magnetic Fields}},  {\em Phys.Rev.} {\bf C81} (2010) 064912,
  [\href{http://xxx.lanl.gov/abs/1003.2436}{{\tt arXiv:1003.2436}}].

\bibitem{Buividovich:2009wi}
P.~V. Buividovich, M.~N. Chernodub, E.~V. Luschevskaya, and M.~I. Polikarpov,
  {\it {Numerical evidence of chiral magnetic effect in lattice gauge theory}},
   {\em Phys. Rev.} {\bf D80} (2009) 054503,
  [\href{http://xxx.lanl.gov/abs/0907.0494}{{\tt arXiv:0907.0494}}].

\bibitem{Buividovich:2010tn}
P.~Buividovich, M.~Chernodub, D.~Kharzeev, T.~Kalaydzhyan, E.~Luschevskaya,
  {\em et.~al.}, {\it {Magnetic-Field-Induced insulator-conductor transition in
  SU(2) quenched lattice gauge theory}},  {\em Phys.Rev.Lett.} {\bf 105} (2010)
  132001, [\href{http://xxx.lanl.gov/abs/1003.2180}{{\tt arXiv:1003.2180}}].

\bibitem{Lifschytz:2009si}
G.~Lifschytz and M.~Lippert, {\it {Anomalous conductivity in holographic QCD}},
   {\em Phys. Rev.} {\bf D80} (2009) 066005,
  [\href{http://xxx.lanl.gov/abs/0904.4772}{{\tt arXiv:0904.4772}}].

\bibitem{Yee:2009vw}
H.-U. Yee, {\it {Holographic Chiral Magnetic Conductivity}},  {\em JHEP} {\bf
  11} (2009) 085, [\href{http://xxx.lanl.gov/abs/0908.4189}{{\tt
  arXiv:0908.4189}}].

\bibitem{Rebhan:2009vc}
A.~Rebhan, A.~Schmitt, and S.~A. Stricker, {\it {Anomalies and the chiral
  magnetic effect in the Sakai- Sugimoto model}},  {\em JHEP} {\bf 01} (2010)
  026, [\href{http://xxx.lanl.gov/abs/0909.4782}{{\tt arXiv:0909.4782}}].

\bibitem{Bardeen:1969md}
W.~A. Bardeen, {\it {Anomalous Ward identities in spinor field theories}},
  {\em Phys. Rev.} {\bf 184} (1969) 1848--1857.

\bibitem{Hill:2006ei}
C.~T. Hill, {\it {Anomalies, Chern-Simons terms and chiral delocalization in
  extra dimensions}},  {\em Phys. Rev.} {\bf D73} (2006) 085001,
  [\href{http://xxx.lanl.gov/abs/hep-th/0601154}{{\tt hep-th/0601154}}].

\bibitem{YeeBNL}
H.~U. Yee, ``{Talk given at Workshop on P- and CP-odd Effects in Hot and Dense
  Matter}.'' Brookhaven, April 26--30, 2010.

\bibitem{Sakai:2004cn}
T.~Sakai and S.~Sugimoto, {\it {Low energy hadron physics in holographic QCD}},
   {\em Prog. Theor. Phys.} {\bf 113} (2005) 843--882,
  [\href{http://xxx.lanl.gov/abs/hep-th/0412141}{{\tt hep-th/0412141}}].

\bibitem{Sakai:2005yt}
T.~Sakai and S.~Sugimoto, {\it {More on a holographic dual of QCD}},  {\em
  Prog. Theor. Phys.} {\bf 114} (2005) 1083--1118,
  [\href{http://xxx.lanl.gov/abs/hep-th/0507073}{{\tt hep-th/0507073}}].

\bibitem{Gorsky:2010xu}
A.~Gorsky, P.~Kopnin, and A.~Zayakin, {\it {On the Chiral Magnetic Effect in
  Soft-Wall AdS/QCD}},  {\em Phys.Rev.} {\bf D83} (2011) 014023,
  [\href{http://xxx.lanl.gov/abs/1003.2293}{{\tt arXiv:1003.2293}}].

\bibitem{Rubakov:2010qi}
V.~A. Rubakov, {\it {On chiral magnetic effect and holography}},
  \href{http://xxx.lanl.gov/abs/1005.1888}{{\tt arXiv:1005.1888}}.

\bibitem{Moore:2010jd}
G.~D. Moore and M.~Tassler, {\it {The Sphaleron Rate in SU(N) Gauge Theory}},
  \href{http://xxx.lanl.gov/abs/1011.1167}{{\tt arXiv:1011.1167}}.

\bibitem{Ghoroku:2007re}
K.~Ghoroku, M.~Ishihara, and A.~Nakamura, {\it {D3/D7 holographic gauge theory
  and chemical potential}},  {\em Phys. Rev.} {\bf D76} (2007) 124006,
  [\href{http://xxx.lanl.gov/abs/0708.3706}{{\tt arXiv:0708.3706}}].

\bibitem{Adler:1969gk}
S.~L. Adler, {\it {Axial vector vertex in spinor electrodynamics}},  {\em Phys.
  Rev.} {\bf 177} (1969) 2426--2438.

\bibitem{Bell:1969ts}
J.~S. Bell and R.~Jackiw, {\it {A PCAC puzzle: $\pi_0 \to \gamma \gamma$ in the
  sigma model}},  {\em Nuovo Cim.} {\bf A60} (1969) 47--61.

\bibitem{Son:2002sd}
D.~T. Son and A.~O. Starinets, {\it {Minkowski-space correlators in AdS/CFT
  correspondence: Recipe and applications}},  {\em JHEP} {\bf 09} (2002) 042,
  [\href{http://xxx.lanl.gov/abs/hep-th/0205051}{{\tt hep-th/0205051}}].

\bibitem{Herzog:2002pc}
C.~P. Herzog and D.~T. Son, {\it {Schwinger-Keldysh propagators from AdS/CFT
  correspondence}},  {\em JHEP} {\bf 03} (2003) 046,
  [\href{http://xxx.lanl.gov/abs/hep-th/0212072}{{\tt hep-th/0212072}}].

\bibitem{Kaminski:2009dh}
M.~Kaminski, K.~Landsteiner, J.~Mas, J.~P. Shock, and J.~Tarrio, {\it
  {Holographic Operator Mixing and Quasinormal Modes on the Brane}},  {\em
  JHEP} {\bf 02} (2010) 021, [\href{http://xxx.lanl.gov/abs/0911.3610}{{\tt
  arXiv:0911.3610}}].

\bibitem{Landsman:1986uw}
N.~P. Landsman and C.~G. van Weert, {\it {Real and Imaginary Time Field Theory
  at Finite Temperature and Density}},  {\em Phys. Rept.} {\bf 145} (1987) 141.

\bibitem{Evans:1995yz}
T.~S. Evans, {\it {The Condensed Matter Limit of Relativistic QFT}},
  \href{http://xxx.lanl.gov/abs/hep-ph/9510298}{{\tt hep-ph/9510298}}.

\end{thebibliography}\endgroup

\end{document}